\documentclass[12pt]{article}
\usepackage[numbers,square,sort]{natbib}
\usepackage[utf8]{inputenc}
\usepackage{amsmath}
\usepackage{graphicx}
\usepackage{amsfonts}
\usepackage[left=2.2cm,top=2.5cm,right=2.2cm]{geometry}
\usepackage{verbatim}
\title{Genetic diversity in introduced populations with Allee effect}
\author{Meike J. Wittmann, Wilfried Gabriel, Dirk Metzler}
\date{Ludwig-Maximilians-Universit\"at M\"unchen, Department Biology II}

\begin{document}
\maketitle

\begin{abstract}
\noindent A phenomenon that strongly influences the demography of small introduced populations and thereby potentially their genetic diversity is the Allee effect, a reduction in population growth rates at small population sizes. We take a stochastic modeling approach to investigate levels of genetic diversity in populations that successfully overcame a strong demographic Allee effect, a scenario in which populations smaller than a certain critical size are expected to decline. Our results indicate that compared to successful populations without Allee effect, successful Allee-effect populations tend to 1) derive from larger founder population sizes and thus have a higher initial amount of genetic variation, 2) spend fewer generations at small population sizes where genetic drift is particularly strong, and 3) spend more time around the critical population size and thus experience more drift there. Altogether, the Allee effect can either increase or decrease genetic diversity, depending on the average founder population size. In the case of multiple introduction events, there is an additional increase in diversity because Allee-effect populations tend to derive from a larger number of introduction events than other populations. Finally, we show that given genetic data from sufficiently many populations, we can statistically infer the critical population size.
\end{abstract}

\vspace{2cm}


\pagebreak

\section{Introduction}
The amount of genetic diversity in a recently established population is strongly shaped by its early history: While the founder population size determines the amount of genetic variation imported from the source population, the population sizes in the following generations influence how much of this variation is maintained and how much is lost through genetic drift. A phenomenon that strongly affects this early history is the demographic Allee effect, a reduction in per-capita growth rate in small populations \citep{2080Stephens1999, 3709Fauvergue2012}.
Allee effects have been detected in species from many different taxonomic groups \citep{302Kramer2009}. Apart from cooperation between individuals, the study subject of the effect's eponym W.C.~Allee \citep{3988Allee1931}, they can result from a variety of other mechanisms such as difficulties to find mating partners, increased predation pressure in small populations, or biased dispersal towards large populations \citep{302Kramer2009}. In this study, we focus on the so-called strong demographic Allee effect, in which the average per-capita growth rate is negative for populations smaller than a certain critical population size \citep{2084Taylor2005}.

A population whose founder size is below this threshold has a high probability of going extinct. With more and more transport of goods around the world, however, many species are introduced to a location not just once, but again and again at different time points. Eventually, a random excess in the number of birth events may cause one of these small introduced populations to grow exceptionally fast, surpass the critical population size, and then grow further to reach high population sizes. Whereas most failed introductions pass unnoticed, the rare successful populations can be detected and sampled and may have substantial impact on native communities and ecosystems. 

Our main question in this study is how expected levels of genetic diversity differ between successful populations that either did or did not have to overcome an Allee effect. Answering this question would help us to understand the ecology and evolution of introduced and invasive populations in several ways. On the one hand, the amount of genetic variation is an indicator for how well an introduced population can adapt to the environmental conditions encountered at the new location. Therefore, the Allee effect---if it influences genetic diversity---could shape the long-term success and impact even of those populations that are successful in overcoming it. On the other hand, genetic patterns created by the Allee effect could help to complete a task that is very challenging when only ecological data are available \citep{3987Courchamp2008,302Kramer2009}: detecting Allee effects in field populations or even estimating the critical population size. Information on the critical population size would be very valuable in practice, for example to identify maximum release rates for species whose establishment is to be prevented, or minimum release rates for those whose establishment is desired, for example in biological control or for species reintroductions \citep{3500Deredec2007}. Furthermore, an important task in statistical population genetics is to reconstruct the demographic history of a population and to infer parameters such as founder population sizes, times since the split of two populations, or migration rates. Should the Allee effect have long-lasting effects on patterns of genetic diversity in established populations, it would have to be taken into account in such analyses.


To our knowledge, there have not been any empirical studies on the population genetic consequences of the Allee effect and the few theory-based results are pointing into different directions. There are arguments suggesting that a strong Allee effect may lead to an increase in genetic diversity, and others that suggest a decrease. An increase in genetic diversity due to the Allee effect is predicted for populations that expand their range in a continuous habitat \citep{605Hallatschek2008,3635Roques2012}. In the absence of an Allee effect, mostly alleles in individuals at the colonization front are propagated. Under an Allee effect, the growth rate of individuals at the low-density front is reduced and more individuals from the bulk of the population get a chance to contribute their alleles to the expanding population. This leads to higher levels of local genetic diversity and weaker spatial genetic structure. A similar effect has been discussed in the spatially discrete case: Kramer and Sarnelle \citep{3699Kramer2008} argued that without Allee effect even the smallest founder populations would be able to grow, leading to populations with very little genetic diversity. The Allee effect, they conclude, sets a lower limit to feasible founder population sizes and thus does not allow for extreme bottlenecks. 

The Allee effect not only influences whether a population will reach high population sizes, but also how fast this happens. So far, the genetic consequences of this change in population dynamics have not been explored theoretically. However, it is often stated that the Allee effect can lead to time lags in population growth \citep{260Drake2006,1007Simberloff2009,867McCormick2010}, i.e. initial population growth rates that are small compared to growth rates attained later \citep{1051Crooks2005}. Such time lags follow almost directly from the definition of the Allee effect and would imply an increased opportunity for genetic drift and thus a reduction in genetic diversity. However, it is not clear whether time lags are still present if we consider the subset of populations that is successful in overcoming the Allee effect. 

In this study, we propose and analyze stochastic models to elucidate and disentangle the various ways in which the Allee effect shapes expected levels of neutral genetic diversity. Furthermore, we investigate under what conditions genetic diversity would overall be lower or higher compared to populations without Allee effect. First, we compare successful populations with and without Allee effect with respect to two aspects of their demography: the distribution of their founder population sizes, i.e. the distribution of founder population sizes conditioned on success, and the subsequent population dynamics, also conditioned on success and meant to include both deterministic and stochastic aspects. In a second step, we will then consider what proportion of neutral genetic variation from the source population is maintained under such a demography. Focusing throughout on introductions to discrete locations rather than spread in a spatially continuous habitat, we first consider the case of a single founding event, and then the case of multiple introductions at different time points. Finally, we explore whether the genetic consequences of the Allee effect could be employed to estimate the critical population size from genetic data.

\section{Model}
\begin{figure}
  \centering
  \includegraphics{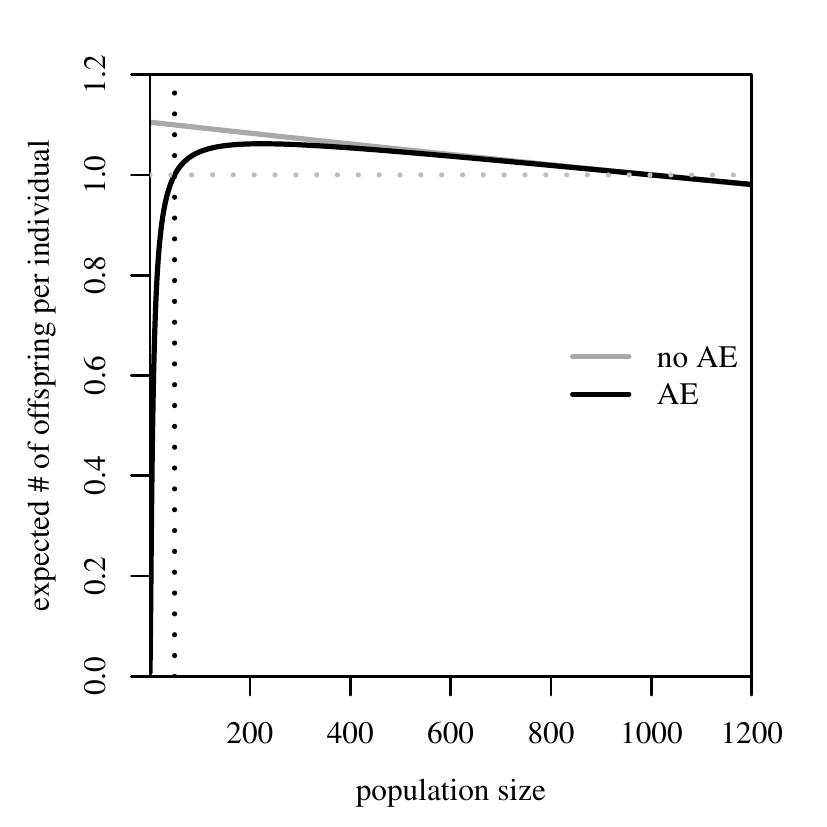}
  \caption{The expected number of surviving offspring per individual ($\lambda(n)$, see equation (\ref{eq:Alleemodel})) as a function of the current population size $n$ without Allee effect (no AE, grey line) or with an Allee effect (AE, black line) of critical size $a=50$ (indicated by dotted vertical line). $k_1=1000, r=0.1$.}
  \label{fig:Alleemodel}
\end{figure}

\renewcommand{\theequation}{2.\arabic{equation}}
\setcounter{equation}{0}
In our scenario of interest, a small founder population of size $N_0$ (drawn from a Poisson distribution) is transferred from a large source population of constant size $k_0$ to a previously uninhabited location. Assuming non-overlapping generations and starting with the founder population at $t=0$, the population size in generation $t+1$ is Poisson-distributed with mean
\begin{equation}
\mathbf{E}[N_{t+1}]=N_t \cdot \lambda(N_t)= N_t \cdot \exp\! \left\{r\cdot \left(1-\frac{N_t}{k_1}\right)\cdot \left(1-\frac{a}{N_t}\right)\right\},
\label{eq:Alleemodel}
\end{equation}
where $k_1$ is the carrying capacity of the new location, $r$ is a growth rate parameter, and $a$ is the critical population size. Unless otherwise noted, we use the parameter values $k_0=10,000$ and $k_1=1000$. To model Allee-effect populations, we set $a=50$, otherwise $a=0$. Under this model, the average per-capita number of surviving offspring per individual $\lambda(N_t)$ is smaller than one for population sizes below the critical population size $a$ and above the carrying capacity $k_1$ and greater than one between critical population size and carrying capacity (figure \ref{fig:Alleemodel}). With $a=0$, this model is a stochastic version of the Ricker model (see e.g. \citep{101Vries2006}). Its deterministic counterpart can exhibit stable oscillations or chaotic behavior for large values of $r$, but here we will only consider values of $r$ between 0 and 2, where $k_1$ is a locally stable fixed point \citep[][p. 29]{101Vries2006}.

We follow the population-size trajectory until the population either goes extinct (unsuccessful population) or reaches target population size $z=100$ (successful population). When a successful population reaches size $z$, we sample $n_s$ individuals from the population and trace their ancestry backwards in time. This allows us to quantify the proportion of genetic variation from the source population that is maintained in the newly founded population. Since the impact of the Allee effect as well as the strength of genetic drift and random population-size fluctuations decline with increasing population size, the particular choice of $z$ and $k_1$ should have little influence on the results as long as they are sufficiently large.

The assumption that each population goes back to a single founding event and then either goes extinct or reaches the target population size $z$ is justified as long as introduction events are rare. Then the fate of a population introduced in one event is usually decided before the respective next event. However, many species are introduced to the same location very frequently \citep{1007Simberloff2009}. Therefore, we also consider a scenario with multiple introduction events: In each generation, an introduction event occurs with probability $p_{intro}$, each time involving $n_{intro}$ individuals. We considered a population successful and sampled it if it had a population size of at least $z$ after the first 200 generations. We fixed the number of generations rather than sampling the population upon reaching $z$ as before, because this would introduce a bias: Populations that would take longer to reach $z$ would be likely to receive more introduction events and thus have higher levels of diversity. With a fixed number of 200 generations and our default choice of migration probability $p_{intro}=0.05$, all populations receive on average ten introduction events. All other parameters were unchanged compared to the case with just one founding event.

\section{Methods}
\renewcommand{\theequation}{3.\arabic{equation}}
\setcounter{equation}{0}
We formulated our demographic model as a Markov chain with transition probabilities
\begin{equation}
P_{ij}=\text{Pr}(N_{t+1}=j|N_t=i)=\frac{e^{-\lambda(i)\cdot i} \cdot (\lambda(i)\cdot i)^j}{j!},
\label{eq:transitionprobs}
\end{equation}
where $\lambda(i)$ is given by equation (\ref{eq:Alleemodel}). We used first-step analysis and Bayes' formula to compute 1) the probability of a population being successful, i.e. reaching some target size $z$ before going extinct, 2) the conditioned distribution of founder population sizes, i.e. the distribution of founder population sizes among successful populations, and 3) the transition probability matrix $\mathbf{P}^c$ of the Markov chain conditioned on reaching $z$ before $0$. The conditioned Markov chain serves two purposes. First, we can use it to directly simulate trajectories of successful populations, which is more efficient than simulating from the original Markov chain and then discarding unsuccessful runs. Second, we can use $\mathbf{P}^c$ to compute the expected number of generations that successful populations with or without Allee effect spend at each of the population sizes from 1 to $z-1$ before reaching $z$, and the expected number of offspring per individual in successful populations with and without Allee effect. Thereby we characterized the population dynamics of successful populations with and without Allee effect. These computations are described in detail in \ref{sec:conditionedMC}. In the case of multiple introduction events, we simulated from the original Markov chain and discarded unsuccessful runs.

Given a successful population size trajectory $N_0,N_1,\dots,N_{T_z}$, we then simulated the genealogies of a sample of $n_s=10$ individuals genotyped at both copies of $n_l=10$ freely recombining loci. We constructed the genealogies by tracing the sampled lineages back to their their most recent common ancestor (see \ref{sec:genealogies} for details). These simulations are based on the assumption that each individual in the offspring generation is formed by drawing two parents independently and with replacement from the parent population. Equivalently, we could assume that each individual is the mother of a Poisson-distributed number of offspring with mean $\lambda(N_t)$ and that the father of each offspring individual is drawn independently and with replacement from the population. Our algorithm for the simulation of genealogies is a discrete-time version of the ancestral recombination graph (see e.g. \citep[][Chapter 7]{4065Griffiths1991,4097Griffiths1997,293Wakeley2009}) with a few modifications to better represent the genetics of very small populations. For each simulation run, we stored the average pairwise coalescence time $G_2$ between sampled chromosomes. To compute the expected proportion of variation from the source population that is maintained in the newly founded population, we divided $G_2$ by $2k_0$, the expected coalescence time for two lineages sampled from the source population. 

We implemented all simulations in C++  \citep{3991Stoustrup1997}, compiled using the g++ compiler \citep[][version 4.7.2]{3992gcc}, and relied on the boost library (version 1.49) for random number generation \citep{3993boost.org2013}. We used R \citep[][version 2.14.1]{575Team2009} for all other numerical computations and for data analysis.

\section{Results}

\subsection{Shift towards larger founder sizes}
\begin{figure}
\begin{center}
  \includegraphics{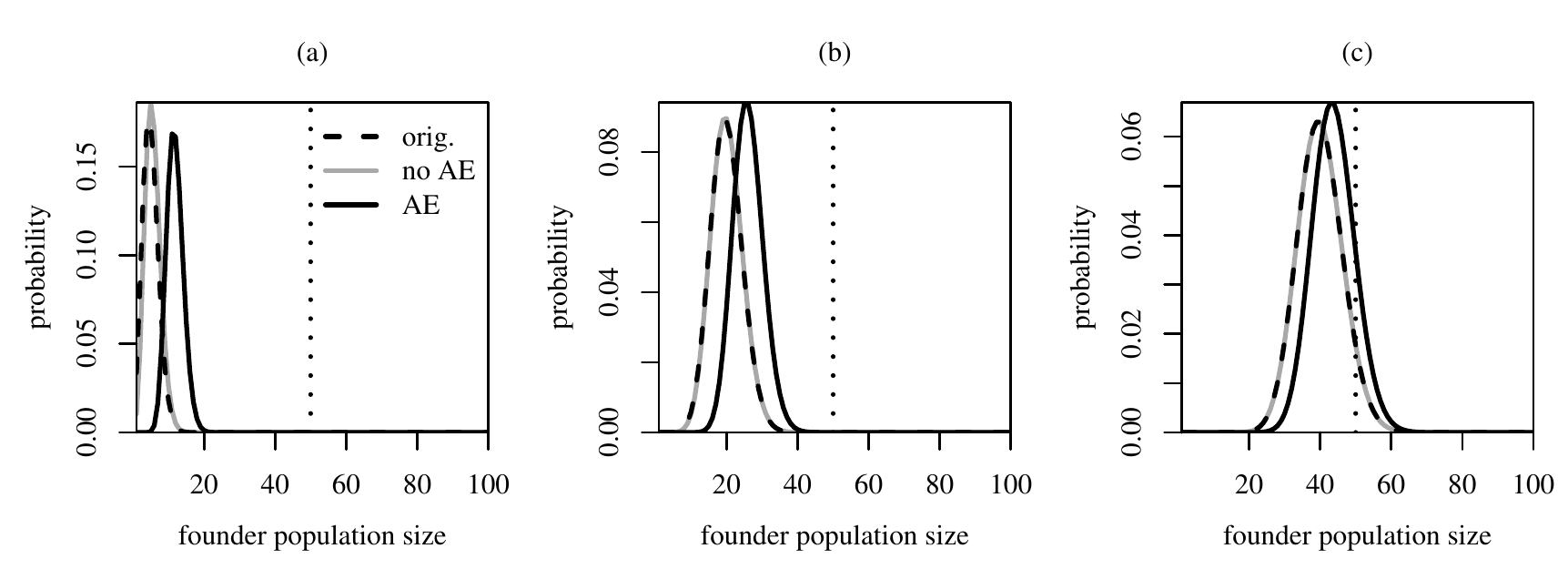}
  \caption{Success-conditioned distributions of founder population sizes with Allee effect (AE, solid black lines) and without (no AE, solid grey lines). The original distribution (dashed) is Poisson with mean 5 (a), 20 (b), or 40 (c) and is almost indistinguishable from the the conditioned distribution without Allee effect in B and C. The dotted vertical line indicates the critical size for Allee-effect populations. Note the differences in the scale of the $y$-axes. $r=0.1$.}
  \label{fig:founderdistributions}
\end{center}
\end{figure}

To compare the demography of successful populations with and without Allee effect, we first examine the distribution of their founder population sizes. These success-conditioned distributions (see \ref{sec:conditionedMC} for how to compute them) differ from the original distribution because the success probability is higher for some founder population sizes than for others. Without Allee effect, small populations can still go extinct by chance, but this quickly becomes very unlikely as the founder population size increases (see \citep{301Dennis2002} and figure \ref{fig:successprob}). Thus, there is a shift towards larger founder population sizes in the conditioned distribution, but this shift is only noticeable for very small average founder sizes (figure \ref{fig:founderdistributions}a). With Allee effect, the success probability is overall lower, even above the critical population size, and has a sigmoid shape with a sharp increase around the critical size (see \citep{301Dennis2002} and figure \ref{fig:successprob}). Consequently, the conditioned distribution of founder population sizes is more strongly shifted to larger population sizes than without Allee effect (figure \ref{fig:founderdistributions}). This shift is particularly strong if the mean of the original distribution is small compared to the critical population size (figure \ref{fig:founderdistributions}a). As the mean founder size approaches the critical population size and a larger proportion of populations is successful (see figure \ref{fig:successprob}), the shift becomes smaller (figure \ref{fig:founderdistributions}b,c).

\subsection{Dynamics of successful populations}
\begin{figure}
  \includegraphics{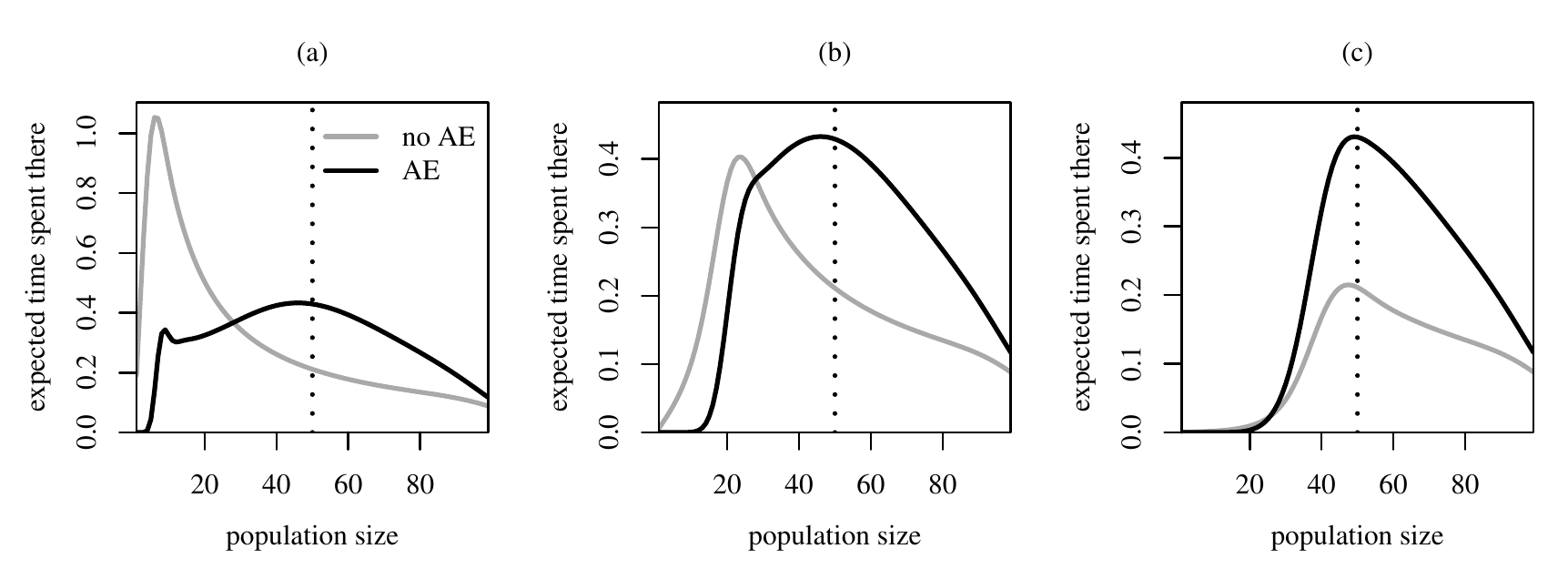}
  \caption{The expected number of generations that successful populations spend at each of the population sizes from 0 to $z-1$ before reaching population size $z$ (here 100). The initial population sizes are 5 in (a), 20 in (b), and 40 in (c). The grey lines represent population dynamics conditioned on success in the absence of an Allee effect (no AE) whereas the black lines represent the conditioned population dynamics with Allee effect (AE) and a critical population size of 50 (indicated by dotted vertical line). Note the differences in the scale of the $y$-axes. $r=0.1$.}
  \label{fig:timespent}
\end{figure}

Upon reaching the target population size $z$, a successful Allee-effect population has on average spent fewer generations at small population sizes than a successful population that did not have to overcome an Allee effect (figure \ref{fig:timespent}), particularly if the founder population size is small compared to the critical population size (figure \ref{fig:timespent}a,b). Thus, although the average Allee-effect population declines at small population sizes (see figure \ref{fig:Alleemodel}), those populations that successfully overcome the critical population size must have grown very fast in this population-size range. Allee-effect populations, however, spend more time at larger population sizes than populations without Allee effect (figure \ref{fig:timespent}). Note that the small peak figure \ref{fig:timespent}a and the kink in figure \ref{fig:timespent}b are due to the fact that the population necessarily spends some time around its founder population size.

\subsection{Population genetic consequences}
\begin{figure}
  \centering
  \includegraphics{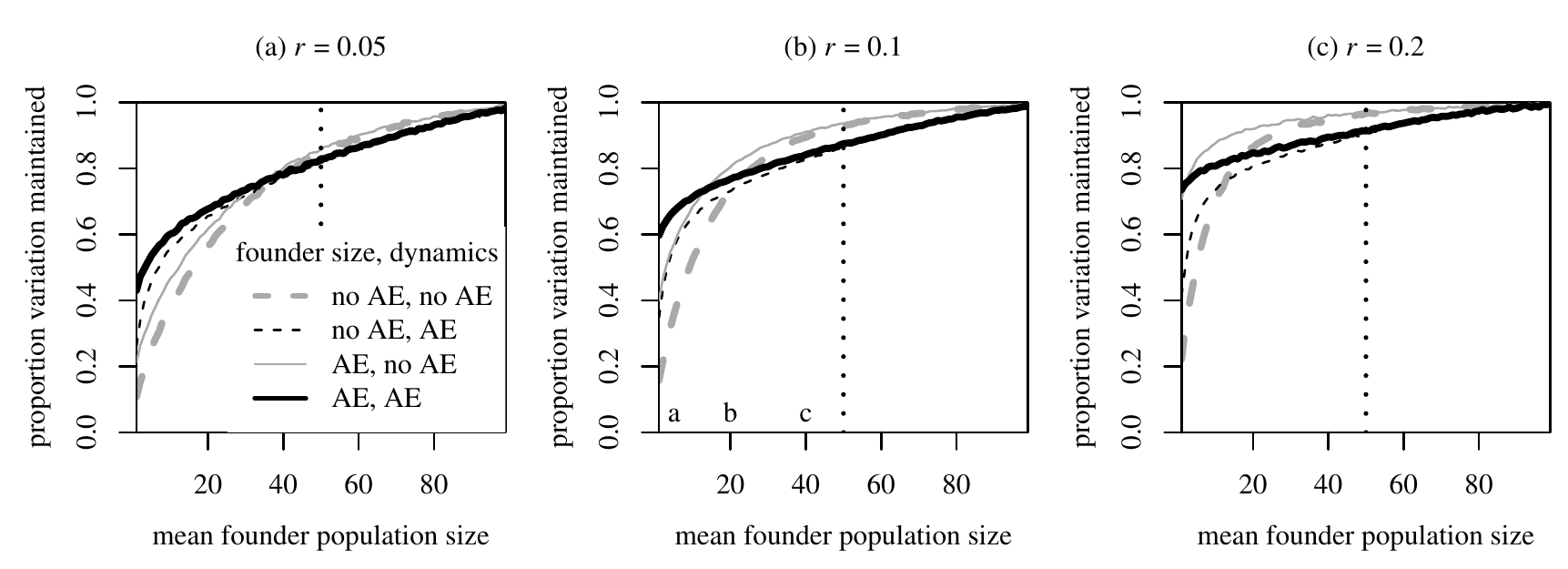}
  \caption{Average proportion of genetic variation from the source population that is maintained by an introduced population upon reaching size $z$. The subplots differ in the value of the growth rate parameter $r$. The values on the $x$-axes correspond to the mean of the original founder-size distribution.  In each subplot, the four displayed scenarios differ in the underlying demography and represent all possible combinations of success-conditioned founder-size distributions either with Allee effect (solid lines) or without (dashed lines) and success-conditioned population dynamics either with Allee effect (black lines) or without (grey lines). Thick lines correspond to the biologically meaningful scenarios with  Allee effect (AE, AE) and without (no AE, no AE), whereas thin lines represent combinations that have no direct biological interpretation but help us to decompose the genetic consequences of the Allee effect (no AE, AE and AE, no AE). The letters a, b, and c in subplot (b) refer to the subplots in figures \ref{fig:founderdistributions} and \ref{fig:timespent}, where we examined for $r=0.1$ and the respective (mean) founder population sizes how the Allee effect influences the conditioned distribution of founder population sizes and the conditioned population dynamics. The critical size for Allee-effect populations ($a=50$) is indicated by a dotted vertical line. Each point represents the average over 20,000 successful populations. Across all points in the plots, standard errors were between 0.0009 and 0.0020, and standard deviations between 0.141 and 0.274.}
  \label{fig:decomposition}
\end{figure}

We have now seen two ways in which the Allee effect modifies the demography of successful populations: it shifts the distribution of founder population sizes and it affects the time they spend in different population-size ranges. In this section, we examine the separate and combined effect of these two features on levels of genetic diversity. Our quantity of interest is the expected proportion of genetic variation from the source population that is maintained by the newly founded population when it reaches size $z$. For different values of the growth rate parameter $r$, we compare four sets of successful populations (figure \ref{fig:decomposition}) representing all possible combinations of a founder-size distribution with or without Allee effect (solid and dashed lines in figure \ref{fig:decomposition}, corresponding to black and grey lines in figure \ref{fig:founderdistributions}, respectively) and subsequent population dynamics with or without Allee effect (black and grey lines in figure \ref{fig:decomposition}, corresponding to  black and grey lines in figure \ref{fig:timespent}, respectively).  

There are three comparisons to be made in each subplot of figure \ref{fig:decomposition}. We first focus on figure \ref{fig:decomposition}b where the growth rate parameter $r$ is the same as in figures \ref{fig:Alleemodel}--\ref{fig:timespent}. We first compare populations with the same dynamics but different distributions of founder population sizes (dashed vs. solid grey lines and dashed vs. solid black lines) and observe that those whose founder population size was drawn from the Allee-effect distribution maintained more genetic variation. This increase was strong for small mean founder population sizes and became weaker with increasing mean founder population size, in accordance with the lessening shift in the conditioned distribution of founder population sizes (see figure \ref{fig:founderdistributions}). Second, among populations that share the founder-size distribution but differ in their population dynamics (black dashed vs. grey dashed lines and black solid vs. grey solid lines), those with Allee-effect dynamics maintained more diversity at small founder population sizes, but less diversity for large founder population sizes. 

Finally, the biologically meaningful comparison is between successful populations with an Allee effect in both aspects of their demography (black solid lines) and successful populations without any Allee effect (grey dashed lines). This comparison reveals the strong and population-size dependent genetic consequences of the Allee effect: For small mean population sizes, successful populations with Allee effect in figure \ref{fig:decomposition}b maintained up to 3.8 times more genetic variation than populations without Allee effect. For mean population sizes close to the critical population size, on the other hand, Allee-populations maintained up to  6.6 \% less genetic variation. Figures \ref{fig:decomposition}a and c show the corresponding results for a smaller and a larger growth rate parameter, respectively. For the smaller growth rate parameter, the Allee effect has a positive effect on genetic diversity over a  wider range of mean founder population sizes (figure \ref{fig:decomposition}a), whereas for a higher growth rate parameter the Allee effect starts to have a negative effect already at relatively small mean founder population sizes (figure \ref{fig:decomposition}c). The results in figure \ref{fig:decomposition} are based on average pairwise coalescence times, a measure related to the average number of pairwise differences in a sample. Results based on the average total length of genealogies were qualitatively similar (see \ref{sec:totallengthstuff}).

\subsection{Multiple introductions}
\begin{figure}
\centering
  \includegraphics{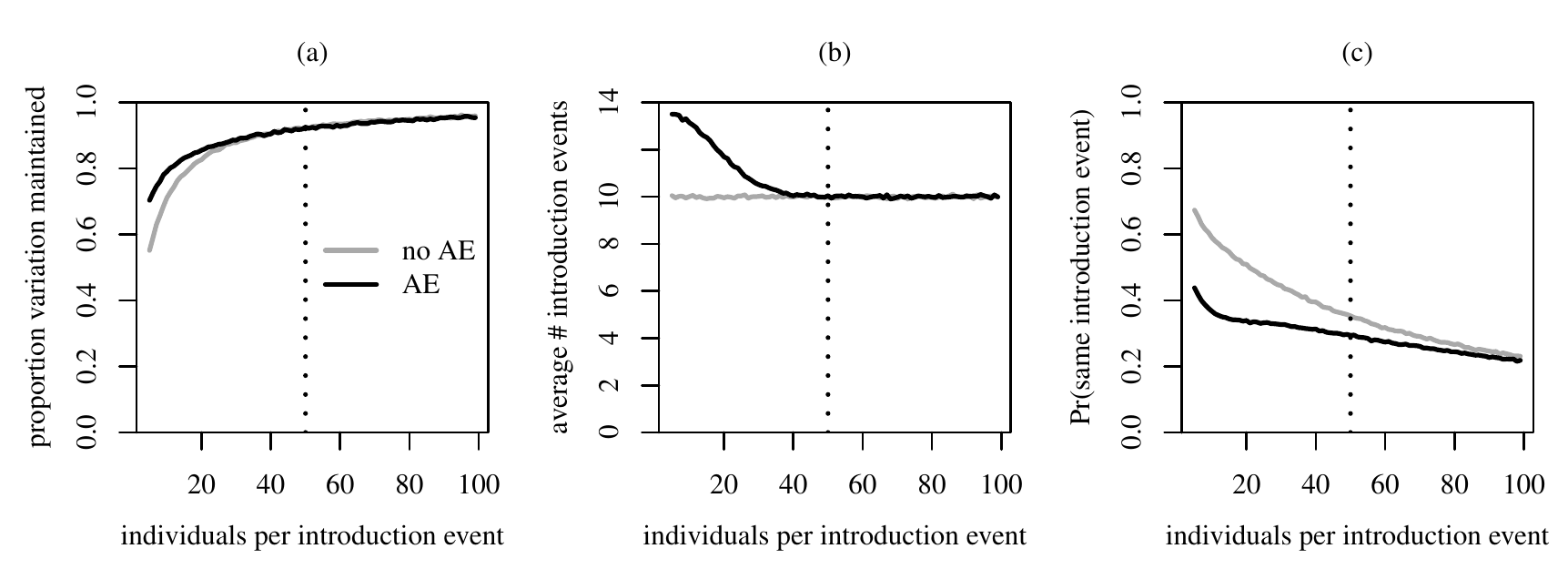}
  \caption{Genetic consequences of the Allee effect in the case of multiple introduction events. (a) Proportion of variation maintained by populations with Allee effect (AE) and without (no AE). Standard deviations were between 0.155 and 0.160 and standard errors between 0.0010 and 0.0012. (b) Average number of introduction events that happened in successful simulation runs. (c) Probability that two lineages in the sample trace back to the same introduction event. Allee-effect populations had a critical population size of 50, as indicated by a dotted vertical line. The migration probability per generation was $0.05$. Each point represents the average over 20,000 successful populations. $r=0.1$.}
  \label{fig:multimig}
\end{figure}

Populations with Allee effect maintained a larger proportion of genetic variation than did populations without Allee effect if the number of individuals introduced per event was smaller than the critical population size (figure \ref{fig:multimig}a). In this parameter range, successful populations with Allee effect had received more introduction events than successful populations without Allee effect (figure \ref{fig:multimig}b). Since in the case of multiple migrations the population can go temporarily extinct, not all introduction events necessarily contribute to the genetic diversity in the sample. However, for small founder population sizes, lineages sampled from an Allee-effect population also had a smaller probability to trace back to the same introduction event than lineages sampled from a population without Allee effect (figure \ref{fig:multimig}c). If a single introduction event was sufficient to overcome the critical population size, there was no noticeable difference between populations with and without Allee effect, neither in the amount of genetic variation maintained nor in the number of introduction events they received. 

\subsection{Estimating the critical population size from genetic data}
\begin{figure}
  \includegraphics{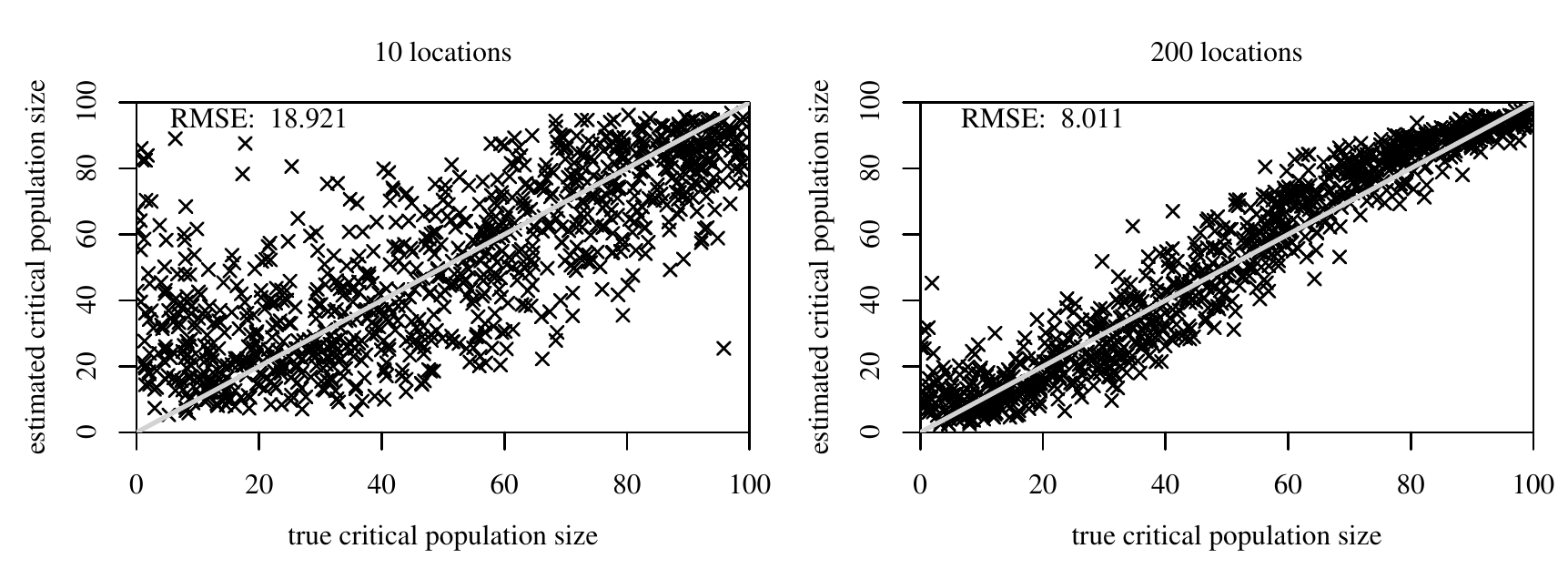}
  \caption{Estimated vs. true values of the critical population size for either 10 or 200 independent locations. On the diagonal grey line, the estimated critical population size is equal to the true one. The value in the upper left corner of each plot is the root mean squared error (RMSE) across the 1000 data sets.}
  \label{fig:ABCresults}
\end{figure}

The results in the last sections have shown that the Allee effect can have substantial impact on the expected amount of genetic variation in a recently founded population. However, due to stochasticity in the population dynamics and genetics, the associated standard deviations are so large that there always is considerable overlap between the underlying distributions with and without Allee effect. Using Approximate Bayesian Computation (ABC), a flexible statistical framework for simulation-based parameter estimation \citep{3457Beaumont2010, 3462Csillery2010} (see \ref{sec:ABCmethods} for the detailed methodology), we explored under what conditions it would be feasible to infer the critical population size from genetic data. We found that it is indeed possible to obtain reasonably accurate estimates of the critical population size, but only if we have information from sufficiently many independent replicates of the process, for example genetic data from several populations that have independently colonized a number of ecologically similar locations (figures \ref{fig:ABCresults} and \ref{fig:RMSE}). 

\section{Discussion}
Our results indicate that the Allee effect strongly influences the expected amount of genetic diversity in a population that recently established from a small founder population size. In the case of a single introduction event, we can attribute this influence to the joint action of three mechanisms: 1) Compared to other successfully established populations, those that have overcome an Allee effect tend to derive from larger founder populations and hence start on average with more genetic diversity. 2) To successfully overcome the critical population size, small Allee-effect populations must grow very fast initially. Therefore, they spend fewer generations in the range of population sizes where genetic drift is strongest, which leads to an increase in genetic diversity relative to populations without Allee effect. 3) Successful Allee-effect populations experience a time lag in population growth around and above the critical population size, leading to increased opportunity for genetic drift and thus a negative effect on genetic diversity. The first and---to some extent---the third mechanism have been suggested before \citep[see][]{3699Kramer2008,867McCormick2010}. In this study, we have clarified the role of the third mechanism and first described the second mechanism in the context of the Allee effect. 

Taken together, the second and third mechanism suggest a peculiar relationship between the original population growth rate and the growth rate among successful populations: Successful populations that are originally expected to decline rapidly (Allee effect-populations substantially below the critical size) grow the fastest, followed by those populations that are expected to increase moderately (populations without Allee effect). The slowest-growing populations are those that are expected to weakly increase or decrease (Allee-effect populations around the critical size). In summary, the per-capita population growth rate conditioned on success (see figure \ref{fig:conditionedAlleemodel}) seems to depend more on the absolute value of the original growth rate, $(E[N_{t+1}]/N_t)-1$, than on its sign, a phenomenon that is also present in simpler models (see \ref{sec:conditioneddiffusion} for an example from diffusion theory). Thus, if we wish to predict the population genetic consequences of the Allee effect, it is not sufficient to know the critical population size, but it may be even more important to determine the absolute value of the average-per capita growth rate at small population sizes.

As the mean founder population size increases, the two mechanisms leading to an increase in diversity (1 and 2) become weaker, whereas the mechanism leading to a decrease in diversity (3) becomes stronger. Therefore, the Allee effect appears to have a positive influence on levels of genetic diversity if typical founder population sizes are small, but a negative effect for large mean founder population sizes. The mean founder population size at which the direction of the effect changes depends on the magnitude of the growth rate parameter $r$. In the case of multiple introduction events, successful populations that have overcome an Allee effect tend to go back to more introduction events than do successful populations without Allee effect, a fourth mechanism that may tip the balance of the genetic consequences of the Allee effect into the positive direction. Exceptionally high levels of genetic diversity caused by Allee effects may contribute to explaining why established alien or invasive populations often harbor a large amount of genetic diversity relative to their source populations \citep{1056Roman2007} although they supposedly established from small numbers. 

As we have seen, the genetic consequences of the Allee effect can be used to estimate the critical population size from genetic data. We conducted our analysis with SNP data in mind, but with different choices of summary statistics other types of genetic data could also be accommodated. To achieve reasonable accuracy, however, we would need independent data from many different locations. Since we found magnitude and direction of the Allee effect's influence to be very context-dependent, it would also be important to know the other demographic parameters fairly well in order to be able to infer the critical population size from genetic data. It could also be worthwhile to perform a joint analysis combining genetic data with relevant ecological information, e.g. on propagule pressure and establishment success \citep{326Leung2004}. As demonstrated by previous studies that addressed other questions in invasion biology with a combination of genetic and ecological data \citep[e.g.][]{3463Estoup2010}, Approximate Bayesian Computation (ABC) provides a flexible statistical framework for such a task.  

Even if it is difficult to detect an Allee effect in genetic data from a single population, neglecting its presence might affect the inference of other demographic parameters such as founder population size, growth rate, and time since the founding event. We explored this possibility in \ref{sec:neglectAllee}, but found no consistent differences in the quality of parameter estimation between populations with and without Allee effect. In both cases, the quality of the inference was rather poor, indicating that the stochastic dynamics in the true model posed a greater challenge to parameter inference than did the Allee effect itself.

Stochastic population models such as the one in this study are not only characterized by their average behavior, but also by the stochastic variability among outcomes. This seems to be particularly important for the genetic consequences of the Allee effect because of several reasons. First, the successful establishment of populations whose size is initially below the critical population size would not be possible in a deterministic model; it requires at least some variability. Second, the extent to which the population dynamics conditioned on success can deviate from the original population dynamics should also depend on the amount of variability. Third, even for a given demographic history, the amount of genetic drift depends on one source of variability, namely that in offspring number among individuals. In this study, we have worked with the standard assumption of Poisson-distributed offspring numbers. However, there is evidence that many natural populations do not conform to this assumption \citep{3977Kendall2010}. Especially in small populations with Allee effect, we would expect more variation in offspring number because many individuals do not encounter a mating partner \citep{302Kramer2009}, whereas those that do can exploit abundant resources and produce a large number of offspring. In a second paper \citep{Secondpaper}, we therefore investigate how the genetic consequences of the Allee effect depend on the distribution of the number of offspring produced by individuals or families. Since the magnitude of the growth rate parameter $r$ affects the relative strength of deterministic and stochastic forces, our results in \citep{Secondpaper} will shed additional light on the role of $r$ for the genetic consequences of the Allee effect.

\section{Acknowledgments}
We would like to thank Raphael Gollnisch and Shankari Subramaniam for assistance with simulations and Pablo Duch\'{e}n for sharing ABC scripts. MJW is grateful for a scholarship from the Studienstiftung des deutschen Volkes.

\pagebreak

\setcounter{section}{0}
\renewcommand{\thesection}{Appendix \arabic{section}}
\renewcommand{\thefigure}{A\arabic{figure}}
\renewcommand{\thetable}{A\arabic{table}}
\renewcommand{\theequation}{A\arabic{equation}}
\setcounter{equation}{0}
\setcounter{figure}{0}

\section{The conditioned Markov chain and its properties}
\label{sec:conditionedMC}
This section explains how to obtain the transition probability matrix of the Markov chain conditioned on the event that the population reaches size $z$ before going extinct (reaching size 0), i.e. conditioned on the event $T_z < T_0$. We also explain how to derive further properties of the conditioned Markov chain. We first restrict our Markov chain to the states $0,1,\dots,z-1,z$, where 0 and $z$ are absorbing states and $1,\dots,z-1$ are transient, that is the Markov chain will leave them at some time. We can write the transition probability matrix of the original Markov chain as
\begin{equation}
\mathbf{P}= \begin{pmatrix}
\mathbf{Q} & \mathbf{R} \\
\mathbf{0} & \mathbf{I}
\end{pmatrix},
\end{equation}
where $\mathbf{Q}$ is a $(z-1) \times (z-1)$ matrix representing the transitions between transient states, $\mathbf{R}$ is a $(z-1) \times 2$ matrix with the transition probabilities from the transient states to the absorbing states $z$ (first column) and 0 (second column), $\mathbf{0}$ is a $2 \times (z-1)$ matrix filled with zeros, and $\mathbf{I}$ is an identity matrix (in this case $2 \times 2$).

Following Pinsky and Karlin \citep{3942Pinsky2010}, we then computed the fundamental matrix $\mathbf{W}=(\mathbf{I}-\mathbf{Q})^{-1}$. $W_{ij}$ gives the expected number of generations a population starting at size $i$ spends at size $j$ before reaching one of the absorbing states. This matrix operation is based on first-step analysis, i.e. on a decomposition of expected quantities according to what happens in the first step (see \citep{3942Pinsky2010} Section 3.4 for details).

\begin{figure}
  \centering
  \includegraphics{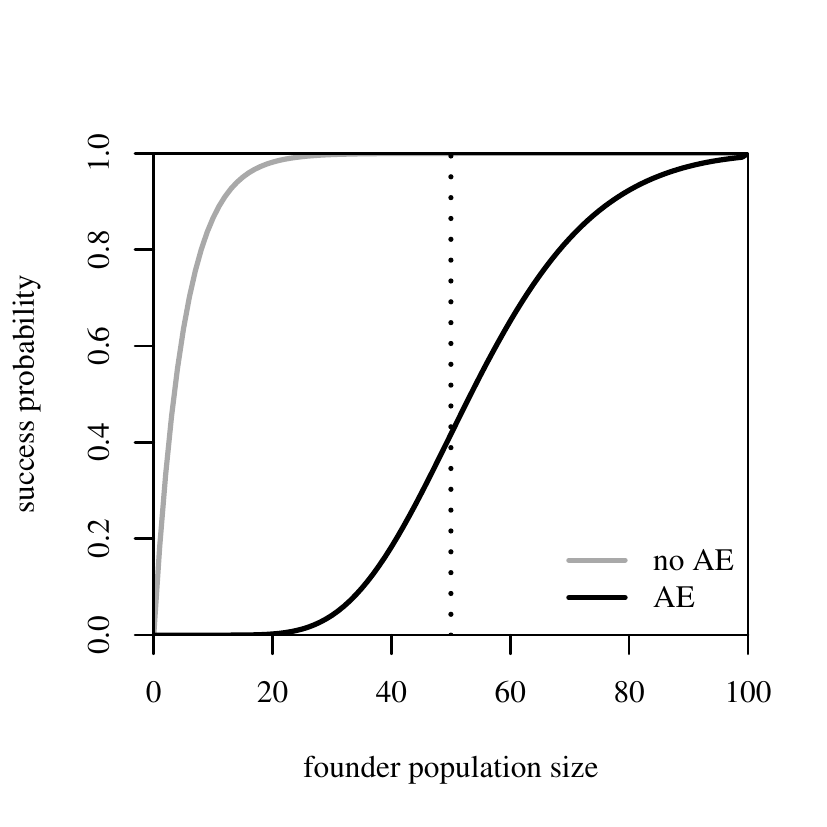}
  \caption{Success probabilities $\text{Pr}(T_{100} < T_0)$ without Allee effect (no AE, grey line) or with an Allee effect (AE, black line) and critical size $a=50$ (indicated by dotted vertical line). $k_1=1000, r=0.1$.}
  \label{fig:successprob}
\end{figure}

The probabilities of absorption in either of the two absorbing states can then be computed as $\mathbf{U}=\mathbf{W}\mathbf{R}$. The first column of $\mathbf{U}$ contains the success probabilities $\text{Pr}(T_z < T_0|N_0=i)$ shown in figure \ref{fig:successprob}. For a given original distribution of founder population sizes (given by the probabilities $\text{Pr}(N_0=n)$ for different founder population sizes $n$), we used the success probabilities together with Bayes' formula to compute the distribution of founder population sizes among successful populations:
\begin{equation}
\text{Pr}(N_0=n|T_z < T_0) = \frac{\text{Pr}(N_0=n)\cdot \text{Pr}(T_z < T_0|N_0=n)}{\sum_{i=1}^{\infty}\text{Pr}(N_0=i)\cdot \text{Pr}(T_z < T_0|N_0=i)}.
\label{eq:conditionedfoundersizes}
\end{equation}
The resulting distributions are shown in figure \ref{fig:founderdistributions}.

Using the success probabilities and Bayes' formula, we then computed the transition probabilities of the Markov chain conditioned on $T_z < T_0$:
\begin{equation}
Q^c_{ij} = \text{Pr}(N_{t+1}=j|N_t=i, T_z < T_0) =\frac{Q_{ij} \cdot \text{Pr}(T_z < T_0|N_0=j)}{\text{Pr}(T_z < T_0|N_0=i)}.
\end{equation}
As $z$ is the only absorbing state of this new Markov chain, the full transition probability matrix is
\begin{equation}
\mathbf{P^c}= \begin{pmatrix}
\mathbf{Q^c} & \mathbf{R^c} \\
\mathbf{0} & 1
\end{pmatrix},
\end{equation}
where $\mathbf{R^c}$ contains the transition probabilities from the transient states to $z$. These probabilities are chosen such that each row sums to 1. In this case, $\mathbf{0}$ stands for a $1 \times (z-1)$ vector filled with zeros. We used this transition probability matrix to simulate the population dynamics conditioned on success.

To further study the conditioned Markov chain, we computed its fundamental matrix $\mathbf{W^c}=(\mathbf{I}-\mathbf{Q^c})^{-1}$. $W^c_{ij}$ gives the number of generations a population starting at size $i$ spends at size $j$ before reaching $z$, conditioned on reaching $z$ before going extinct. These are the values shown in figure \ref{fig:timespent}. Note that in these plots we did not include the first generation, which the population necessarily spends at its founder size. 

\begin{figure}
  \centering
  \includegraphics{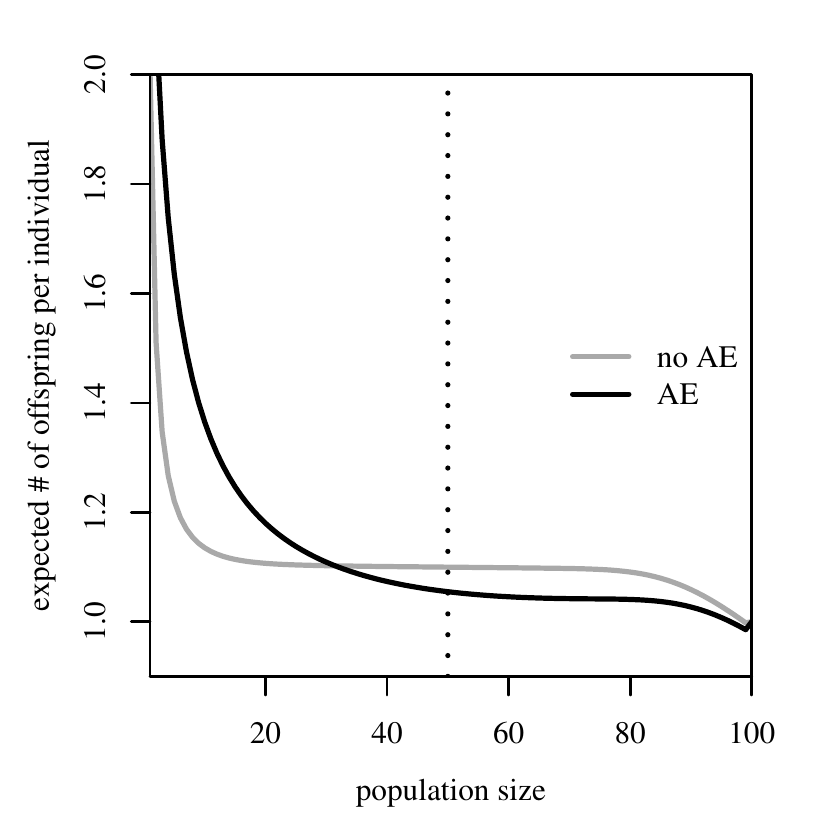}
  \caption{The expected number of surviving offspring per individual in successful populations (see equation \eqref{eq:conditionedpcgr}) as a function of the current population size without (no AE, grey line) or with an Allee effect (AE, black line) and critical size $a=50$ (indicated by dotted vertical line). $k_1=1000, r=0.1$.}
  \label{fig:conditionedAlleemodel}
\end{figure}

We also computed the expected number of surviving offspring per individual at population size $i$ under the conditioned population dynamics (figure \ref{fig:conditionedAlleemodel}):
\begin{equation}
\frac{1}{i}\sum_{j=1}^z j \cdot P^c_{ij}.
\label{eq:conditionedpcgr}
\end{equation}
This is an approximation because our Markov chain is restricted to population sizes up to $z$ whereas actual populations would be able to grow beyond $z$. However, in the range of population sizes that is most relevant for our study, i.e. at small and intermediate population sizes, equation \eqref{eq:conditionedpcgr} should give an accurate approximation of the expected number of surviving offspring per individual.

\pagebreak

\section{Details on the simulation of genealogies}
\label{sec:genealogies}
The genealogies are constructed by tracing the ancestry of the sampled genetic material backwards in time. One special feature of our algorithm is the possibility of multiple and simultaneous mergers. This means that in one generation several coalescent events can happen and each of them can possibly involve more than two lineages. Such events are very rare in large populations as assumed by standard coalescent models but they can be quite frequent in small populations as we are considering here. Another special feature is that the genealogical process takes into account explicitly that individuals are diploid and bi-parental and thus avoids logical inconsistencies that may occur when independently simulating the genealogies at different loci \citep{3718Wakeley2012}. However, this realism comes at a computational cost and in cases where we are only interested in average levels of genetic diversity, i.e. for the analysis underlying figures \ref{fig:decomposition}, \ref{fig:multimig}, and \ref{fig:decomposition_Ttotal} we resorted to independently simulating the genealogies at the different loci.

The current state of the ancestry is defined by a set of lineage packages for each population (source population and newly founded population). Such a lineage package contains all the genetic material that is traveling within the same individual at that time point. It has two sets of slots, one set for each genome copy. Each set has a slot for each locus. If the genetic material at a certain locus and genome copy is ancestral to the sample, the slot is occupied by a node, otherwise it is empty.

\begin{figure}
  \centering
  \includegraphics[width=\textwidth]{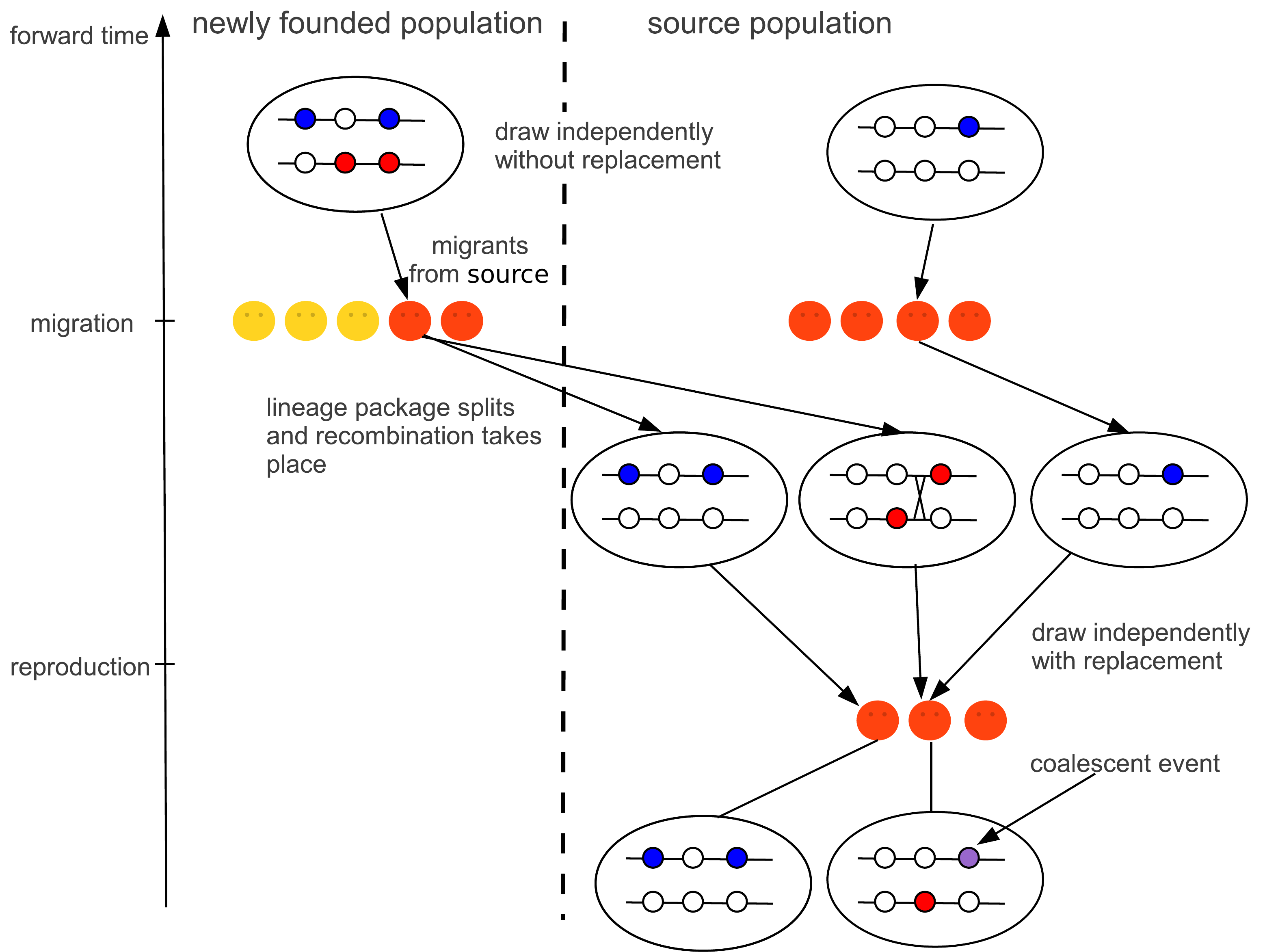} 
  \caption{Illustration of the backward-in-time simulation of genealogies.}
  \label{fig:genealogyillustration}
\end{figure}

The ancestral history starts with $2 \cdot n_s$ lineage packages in the newly founded population. Initially all slots in the lineage packages are occupied by nodes. 
From there, the ancestry is modeled backwards in time until at each locus there is just one node left. Given the state of the ancestry in generation $t$, the state in generation $t-1$ is generated as follows: Backward in time, each generation starts with a migration phase (figure\ \ref{fig:genealogyillustration}). All lineage packages that are currently in the newly founded population choose uniformly without replacement one of the $Y_t$ migrants from the source population, or one of the $N_t - Y_t$ residents. Note that in our simulations with a single founding event, $Y_0 = N_0$ and $Y_t = 0$ for all $t>0$, whereas in the case of multiple introductions, $Y_t$ can be positive also at $t>0$. According to their choice in this step, lineage packages either remain in the newly founded population or are transferred to the source population.

Then each lineage package splits into two because the two genome copies (sets of slots) each derive from a possibly different parent (see figure\ \ref{fig:genealogyillustration}). Lineage packages that do not contain ancestral material are discarded immediately. For each of the remaining lineage packages, recombination is implemented by independently constructing a stochastic map $R: \{1,\dots,n_{l}\} \to \{0,1\}$ such that
\begin{equation}
R(1)=\begin{cases}
0 & \text{with probability } \frac{1}{2}\\
1 & \text{with probability } \frac{1}{2}
\end{cases}
\end{equation}
and then
\begin{equation}
R(n+1)=\begin{cases}
R(n) & \text{with probability } 1-\rho_n \\
1-R(n) & \text{with probability } \rho_n \\
\end{cases}
\end{equation}
is drawn recursively for $n \in \{1,\dots,n_{l}-1\}$. The recombination probability $\rho_n$ between loci $n$ and $n+1$ was 0.5 for all analyses in this study. A node at locus $l$ in the new lineage package is placed into the first genome copy if $R(l)=0$ and into the second genome copy if $R(l)=1$.

After each lineage package underwent splitting and recombination, all resulting lineage packages uniformly pick one of the $N_{t-1}$ or $k_0$ individuals as ancestor, depending on whether they are in the newly founded or in the source population, this time with replacement (see figure\ \ref{fig:genealogyillustration}). Lineage packages that chose the same ancestor are merged. If there is more than one node at the same genome copy and slot, a coalescent event takes place.

Because genetic drift is strong in small populations, many pairs of lineages will already encounter their common ancestor within the newly founded population. The lineages that did not coalesce until time 0 must all be in the source population which is assumed to be of constant size $k_0$ at all times. To efficiently simulate the genealogical process of the remaining lineages, we follow one of two procedures: As long as there is still lineage packages that carry more than one node or if the number of pairs of remaining lineages is larger than $k_0/10$, we continue as before, going backwards generation-by-generation. Each lineage package can split due to recombination and merge with others that choose the same ancestor. However, as the source population is large it would take a long time until all lineages find their most recent common ancestor (MRCA) and in most generations nothing would happen. Furthermore, nodes within the same lineage package typically become separated by recombination relatively fast. Thus whenever there is no lineage package with more than one node, we switch to a second and more efficient simulation mode: If $n_{total}$ is the number of lineage packages, we draw the number of generations $T$ until the next merger of two lineage packages from a geometric distribution with success probability
\begin{equation}
p_{merge}=\frac{\binom{n_{total}}{2}}{k_0}
\end{equation}
and update the current time to $t-T$. Then one of the $\binom{n_{total}}{2}$ pairs of lineage packages is picked at random and the lineage packages are merged. If the two lineage packages have their node at the same slot, a coalescent event happens and we continue in the efficient simulation mode. If the nodes are at different slots we again have a lineage package with more than one node and we switch back to the more accurate simulation mode. Note that the last recombination event in the accurate simulation mode ensures that each node has a 50 \% chance to be in the first or second genome copy. This leads to a 50 \% chance of coalescence if two lineage packages with a node at the same locus merge. Thus, there is no need to implement recombination in the efficient simulation mode. This efficient simulation mode excludes multiple and simultaneous mergers, events that should be very rare for a reasonably large source population size $k_0$.

We switch between the two simulation modes until eventually there is only one node left at each locus, the MRCA of all sampled genetic material at the respective locus. Throughout the simulation, we store all information needed to provide the topology and branch lengths (in number of generations) for the genealogies at each locus.

\section{Results based on total length of the genealogy}
\label{sec:totallengthstuff}
\begin{figure}[h]
  \centering
  \includegraphics{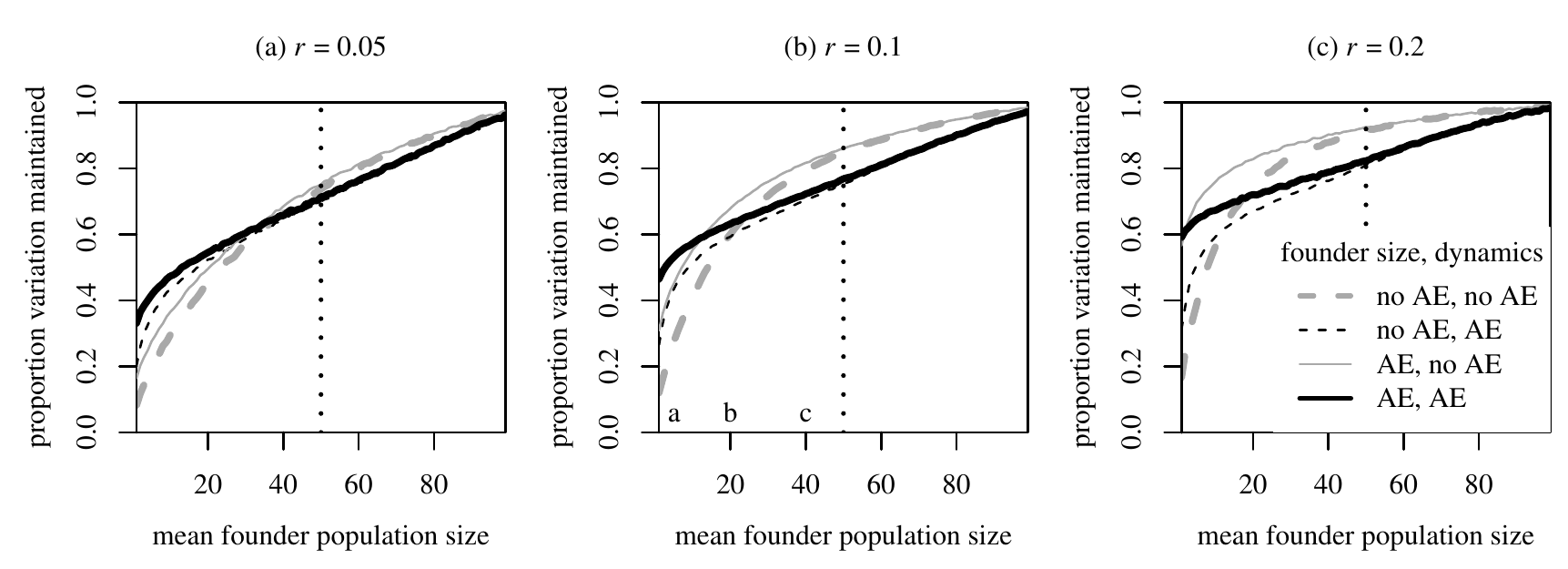}
  \caption{Average proportion of genetic variation maintained (based on the average total length of sample genealogies) as a function of the mean of the original (unconditioned) founder population size distribution. The four scenarios differ in the underlying demography and represent all possible combinations of success-conditioned founder-size distributions either with Allee effect (solid lines) or without (dashed lines) and success-conditioned population dynamics either with Allee effect (black lines) or without (grey lines). Thick lines correspond to the biologically meaningful scenarios with  Allee effect (AE, AE) and without (no AE, no AE), whereas thin lines represent combinations that have no direct biological interpretation but help us to decompose the genetic consequences of the Allee effect (no AE, AE and AE, no AE). The letters a, b, and c in subplot (b) refer to the subplots in figures \ref{fig:founderdistributions} and \ref{fig:timespent}, where we examined for the respective (mean) founder population sizes how the Allee effect influences the conditioned distribution of founder population sizes and the conditioned population dynamics. The critical size for Allee-effect populations ($a=50$) is indicated by a dotted vertical line. Each point represents the average over 20,000 successful populations. Across all points, standard errors were between 0.0007 and 0.0016, and the corresponding standard deviations between 0.103 and 0.221.}
  \label{fig:decomposition_Ttotal}
\end{figure}

In the main text, we use average pairwise coalescence times to assess genetic diversity. Here we show the corresponding results for the average total length of the genealogy $G_{total}$, a measure related to the number of segregating sites or the number of alleles in a sample. To measure the proportion of variation maintained, we divided $G_{total}$ by $4k_0 \cdot \sum_{i=1}^{2n_s-1}\frac{1}{i}$, the expected total length of the sample genealogy if all lineages would have been sampled in the source population \citep[][p. 76]{293Wakeley2009}. The results (figure \ref{fig:decomposition_Ttotal}) were qualitatively similar to the results based on average pairwise coalescence times (see figure \ref{fig:decomposition}), except that the proportion of variation maintained more slowly approached one with increasing founder population size.

\section{Methodology for estimating the critical population size}
\label{sec:ABCmethods}
We generated 1000 pseudo-observed data sets and 100,000 simulated data sets, each with independent introductions to 200 locations. The critical population sizes were drawn from a uniform distribution on [0,100]. We fixed the other parameters of the population dynamics ($k_0=10,000, k_1 = 1000, r=0.1$) and assumed them to be known with certainty. We further assumed that the original distribution of founder population sizes was Poisson with mean 20, and sampled the founder population sizes independently for each location from the conditioned distribution of founder population sizes for the respective critical population size. Given the selected founder population size, we simulated the population dynamics at each location from the conditioned Markov chain until the population reached size 200, i.e. twice the largest possible critical population size. 

At this point, we sampled $n_l=10$ individuals at both genome copies, resulting in a sample of 20 chromosomes from a given location. We generated genealogies for 10 freely recombining loci. To obtain a more differentiated picture of patterns of genetic variation and capture as much information as possible, we did not use the average pairwise coalescence times or total lengths of the genealogy as before. Instead, for $i \in 1,2,\dots,19$ we took the combined length of all branches $B_i$ that have $i$ descendants in the sample. Using these branch lengths and assuming that the number of mutations on a branch of length $b$ is Poisson-distributed with parameter $\mu \cdot b$, we estimated the mean and variance across loci of the entries of the site-frequency spectrum (SFS) $\xi_i$, i.e. the number of mutations that appear in $i$ chromosomes in the sample, as
\begin{equation}
\hat{\mathbf{E}}[\xi_i]=\mu \cdot \bar{B_i}
\label{eq:estimatedmean}
\end{equation}
and, using the law of total variance,
\begin{equation}
\hat{\mathbf{Var}}[\xi_i]= \hat{\mathbf{E}}\left[\mathbf{Var}[\xi_i | B_i]\right]+ \hat{\mathbf{Var}}\left[\mathbf{E}[\xi_i | B_i]\right]=\mu \cdot \bar{B_i} + \mu^2 \cdot s^2(B_i),
\label{eq:estimatedvariance}
\end{equation}
where the $\bar{B_i}$ are the average branch lengths across the $n_l$ loci and the $s^2(B_i)$ are the corresponding empirical variances. We assumed $\mu=0.001$. Note that we do not take into account variability introduced by the mutation process because we assume that we have enough loci to estimate the means and variances of the SFS entries with reasonable accuracy.

\begin{figure}
  \centering
  \includegraphics{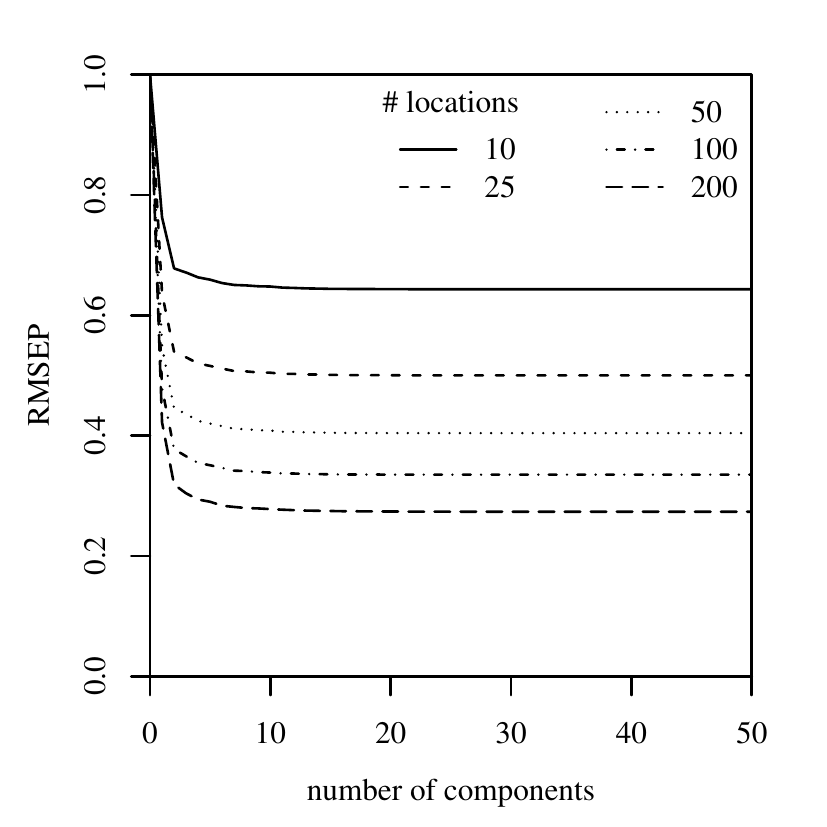}
  \caption{Root mean squared error of prediction (RMSEP) as a function of the number of PLS components for various numbers of locations. In no case did the RMSEP substantially change beyond 20 components. Thus, we decided to include 20 components as summary statistics for ABC.}
  \label{fig:componentselection}
\end{figure}

We further summarized the data for each SFS entry $i \in 1,2,\dots,19$ by computing the averages and empirical standard deviations of the quantities in eqs. \eqref{eq:estimatedmean} and \eqref{eq:estimatedvariance} across locations. To investigate how the quality of the estimation depends on the number of independent locations available, we took into account either only 10, 25, 50, 100, or all 200 of them to compute these statistics. Using the pls script from abctoolbox \citep{3941Wegmann2010} and the pls package in R \citep{3678Mevik2007}, we then conducted partial least squares regression on the first 10,000 simulated data sets to condense the information contained in the 76 summary statistics to a smaller number of components. To decide on the number of components, we examined plots of the root mean squared error of prediction (RMSEP) as a function of the number of components (figure \ref{fig:componentselection}). For none of the different numbers of locations did the RMSEP change substantially beyond 20 components. Thus, we decided to include 20 components as summary statistics for ABC.

\begin{figure}
  \centering
  \includegraphics{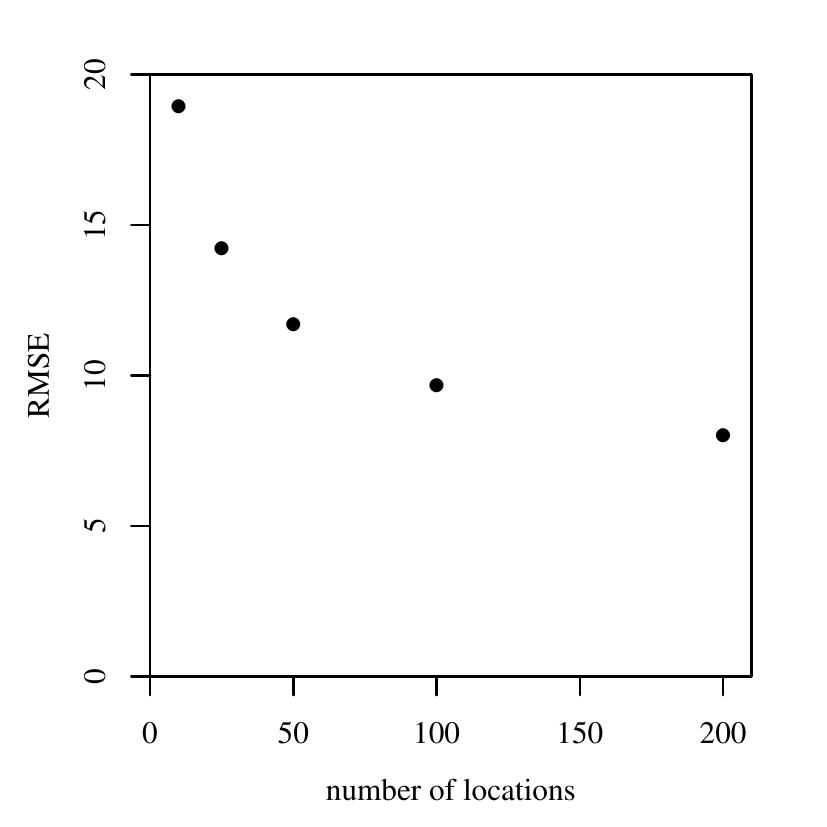}
  \caption{The root mean squared error (RMSE) of the estimated critical population size  as a function of the number of independent locations used for the estimation.}
  \label{fig:RMSE}
\end{figure}

\begin{figure}
  \centering
  \includegraphics{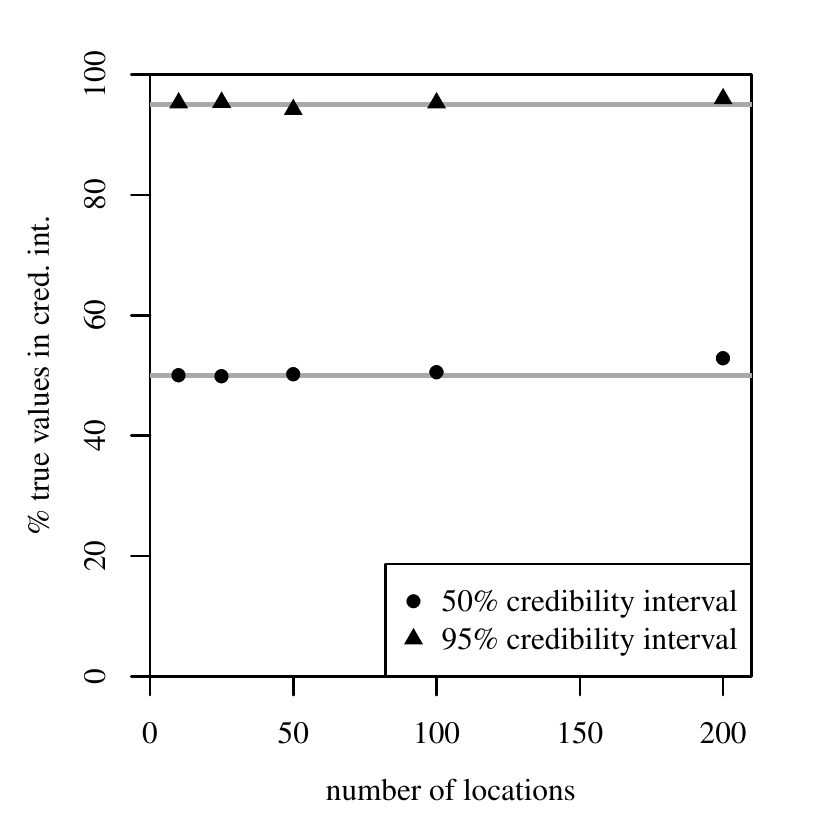}
  \caption{Percentage of true parameter values that fall within the 50\% and 95\% credibility interval, an indicator for how well Approximate Bayesian Computation approximates Bayesian inference. The grey lines are at 50\% and 95\%.}
\label{fig:credibility}
\end{figure}

We used these 20 PLS components as summary statistics for parameter estimation with the R package abc \citep{3940Csillery2012}. We chose a tolerance of 1 \% and used the option ``loclinear'' implementing the local linear regression method \cite{657Beaumont2002}. To avoid estimated parameter values that fall outside the prior, we estimated $\ln(a/(100-a))$ and then back-transformed the estimated values. For each pseudo-observed data set, we thus used the 100,000 simulated data sets to approximate the posterior distribution of the critical population size given a uniform prior on [0,100]. For each data set, we stored the mean of the posterior, which we take as our point estimator, and the 50 \% and 95 \% credibility intervals. We observed that the quality of parameter inference improved with an increasing number of locations (figures \ref{fig:ABCresults} and \ref{fig:RMSE}). An examination of the percentage of pseudo-observed data sets for which the true parameter value falls into the respective 50 \% or 95 \% credibility interval suggests that ABC approximates Bayesian inference reasonably well in this case (figure \ref{fig:credibility}).

\pagebreak

\section{Consequences of a neglected Allee effect}
\label{sec:neglectAllee}
Using the ABC framework again, we explored the consequences of neglecting the Allee effect when estimating other demographic parameters: the founder population size $N_0$, the growth parameter $r$, and the number of generations since the founding event. We generated 2000 pseudo-observed data sets from our stochastic model, 1000 without Allee effect and 1000 with an Allee effect and a critical population size of 50. As the basis for estimation in ABC, we used 100,000 data sets that were simulated from a model without Allee effect. To also explore the consequences of neglecting stochasticity, we considered two versions of the model without Allee effect: our stochastic model with $a=0$ and a modified version where we removed as much stochasticity as possible. That is, the population size in the next generation was not drawn from a Poisson distribution, but was set to $\mathbf{E}[N_{t+1}]$ if this value was an integer. Otherwise, we randomly set $N_{t+1}$  to the next smallest or next largest integer with the respective probabilities chosen such that equation \eqref{eq:Alleemodel} was fulfilled.

The priors for the demographic parameters of interest were as follows: 
\begin{equation}
\ln(N_0)\sim \text{unif}([\ln(5),\ln(80)]),
\end{equation}
\begin{equation}
r \sim \text{unif}([0.01,0.1])
\end{equation}
and 
\begin{equation}
n_g \sim \text{unif}(\{20,\dots,500\}), 
\end{equation}
where unif stands for the uniform distribution. The other parameters were fixed: $k_0=10,000, k_1=1000, \mu=0.001, n_s=10$. For each data set, we retried simulating with the same parameter combination until we obtained a successful population with $N_{n_g}\geq n_s$. We generated 100 independent genealogies for samples of size $n_s$ taken at time $n_g$ and computed means and variances of the entries of the site-frequency spectrum as described in \ref{sec:ABCmethods}. Using partial least squares regression on the first 10,000 simulated data sets, we reduced this information to 20 components that served as summary statistics for ABC. As above, we used the R package abc \citep{3940Csillery2012} with a tolerance of 1 \% and the option ``loclinear'. 

\begin{figure}
  \centering
  \includegraphics{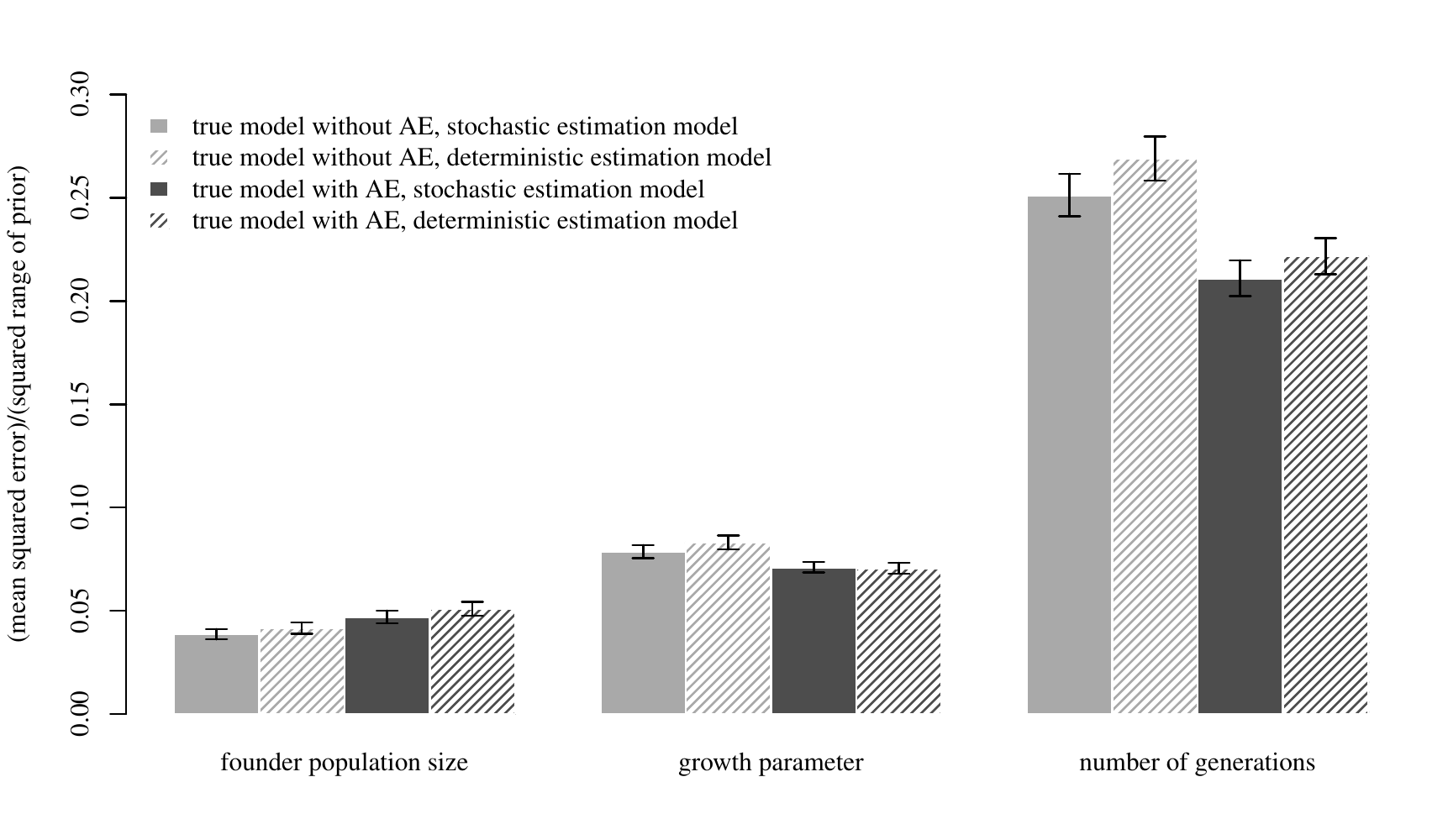}
  \caption{Mean squared error (MSE) $\pm$ its standard error, relative to the squared range of the prior for different demographic parameters in an Approximate Bayesian Computation analysis that neglects the Allee effect.}
  \label{fig:ignorantABCresults}
\end{figure}

In figure \ref{fig:ignorantABCresults}, we compare the quality of parameter estimation across the four possible combinations of whether or not the true model includes an Allee effect and whether the model used for estimation was stochastic or deterministic. The differences in quality between the four combinations were not consistent across estimated parameters. Overall, the quality of the estimation was poor, with root mean squared errors of up to half the range of the corresponding prior. Note that these problems cannot only result from model misspecification since the case where the correct model was used (solid light grey bars in figure \ref{fig:ignorantABCresults}) also produced large errors. Thus, it appears that the amount of stochasticity in the model is so large as to prevent accurate parameter inference based on genetic data from a single population.

\pagebreak

\section{A conditioned diffusion process}
\label{sec:conditioneddiffusion}
Our results in the main text indicate that the growth rate under the conditioned population dynamics depends mostly on the absolute value and not so much on the sign of the growth rate under the unconditioned population dynamics. For mathematically interested readers, we now explore a simple model where this fact can be proven easily. We consider a diffusion process on the interval $[0,1]$ with constant infinitesimal mean $\mu(x)=\mu$ and constant infinitesimal variance $\sigma^2(x)=\sigma^2$. We will show that the associated diffusion process conditioned on hitting 1 before 0 is independent of the sign of $\mu$ and that its infinitesimal mean increases with $|\mu|$.

\begin{figure}
\centering
\includegraphics{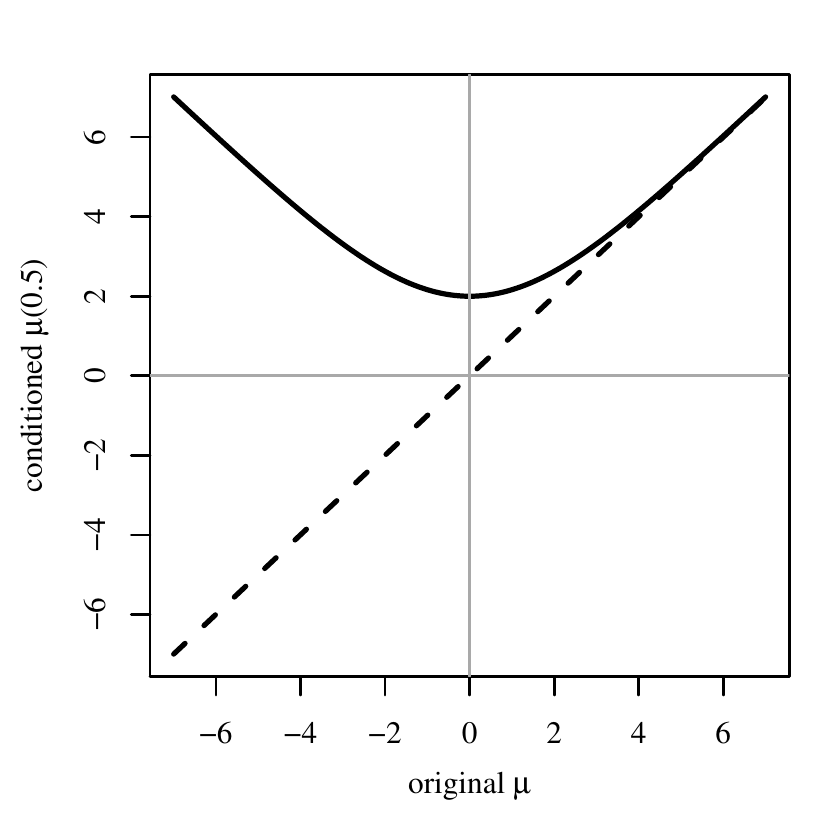}
\caption{The infinitesimal mean of the conditioned diffusion process at 0.5 (solid line), i.e. in the middle of the interval, as a function of the infinitesimal mean of the original process. On the dashed line, the infinitesimal means of original and conditioned process would be equal.}
\label{fig:mustar}
\end{figure}

Our task is to compute the infinitesimal mean and variance of the conditioned diffusion process. Following the formulas given by Karlin and Taylor \citep[][p. 263]{3238Karlin1981}, the infinitesimal mean of the conditioned diffusion process is
\begin{equation}
\mu^*(x)=\mu(x)+\frac{s(x)}{S(x)}\cdot \sigma^2(x),
\label{eq:muprime}
\end{equation}
where $S(x)$ is the scale function and $s(x)$ is its derivative. Using the definitions of these functions \citep[e.g.][p. 262]{3238Karlin1981} and plugging in the parameters of our diffusion, we obtain
\begin{equation}
s(x)= \exp\!\left(-\int_0^x \frac{2\mu(\eta)}{\sigma^2(\eta)} d\eta\right)=\exp\!\left(-\frac{2\mu x}{\sigma^2}\right)
\label{eq:smalls}
\end{equation}
and
\begin{equation}
S(x)=\int_0^x s(\eta) d\eta = \frac{\sigma^2}{2\mu} \cdot \left[1-\exp\!\left(-\frac{2\mu x}{\sigma^2}\right)\right].
\label{eq:capitalS}
\end{equation}
Substituting eqs. \eqref{eq:smalls} and \eqref{eq:capitalS} into equation \eqref{eq:muprime}, we obtain
\begin{equation}
\mu^*(x)=\mu \cdot \frac{\exp(2\mu x/\sigma^2) +1}{\exp(2\mu x/\sigma^2) -1}=:f(\mu),
\end{equation}
a function that is symmetric about 0, i.e. $f(-\mu)=f(\mu)$, and increases with the absolute value of $\mu$ (figure \ref{fig:mustar}). The variance $\sigma^{2*}(x)$ equals the original variance $\sigma^2(x)$.

\pagebreak


\end{document}